\documentstyle[12pt,preprint,aps,floats]{revtex}

\tighten

\input epsf.tex

\newcommand{\postscript}[2]{\setlength{\epsfxsize}{#2\hsize}
   \centerline{\epsfbox{#1}}}
\newcommand{\sign}{\:\!\text{sign}\:\!}
\newcommand{\mgaugino}{M_{1/2}}
\newcommand{\mt}{m_t}
\newcommand{\mweak}{M_{\text{Weak}}}
\newcommand{\msusy}{M_{\text{SUSY}}}
\newcommand{\mgut}{M_{\text{GUT}}}

\newcommand{\tb}{\tan\beta}
\newcommand{\bold}[1]{{\text{\normalsize\boldmath $#1$}}}

\newcommand{\gev}{\text{GeV}}
\newcommand{\tev}{\text{TeV}}
\newcommand{\cm}{\text{cm}}

\newcommand{\Omegachi}{\Omega_{\chi}}
\newcommand{\be}{\begin{equation}}
\newcommand{\ee}{\end{equation}}

\newcommand{\bsg}{B\to X_s \gamma}

\newcommand{\thetacp}{\theta_{\text{CP}}}

\begin{document}

\draft

\renewcommand{\thefootnote}{\fnsymbol{footnote}}
\setcounter{footnote}{0}

\preprint{
\noindent
\hfill
\begin{minipage}[t]{3in}
\begin{flushright}
MIT--CTP--3032\\
CERN--TH/2000--305\\
IASSNS--HEP--00--60\\
FERMILAB--PUB--00/202--T\\
hep-ph/0011356
\end{flushright}
\end{minipage}
}

\title{
\vskip 0.3in
Focus Point Supersymmetry: Proton Decay,\\
Flavor and CP Violation, and the Higgs Boson Mass}

\author{
Jonathan L.~Feng$^a$\footnote{E-mail: jlf@mit.edu} and Konstantin
T.~Matchev$^b$\footnote{E-mail: Konstantin.Matchev@cern.ch} 
\vskip 0.2in
}

\address{
  ${}^{a}$
  Center for Theoretical Physics,
  Massachusetts Institute of Technology\\
  Cambridge, MA 02139, U.S.A.\\ 
\vskip 0.1in
  ${}^{b}$
  Theory Division, CERN, CH--1211
  Geneva 23, Switzerland}

\maketitle

\begin{abstract}

In focus point supersymmetry, all squarks and sleptons, including
those of the third generation, have multi-TeV masses without
sacrificing naturalness.  We examine the implications of this
framework for low energy constraints and the light Higgs boson mass.
Undesirable contributions to proton decay and electric dipole moments,
generic in many supersymmetric models, are strongly suppressed.  As a
result, the prediction for $\alpha_s$ in simple grand unified theories
is 3$\sigma$--5$\sigma$ closer to the experimental value, and the
allowed CP-violating phases are larger by one to two orders of
magnitude.  In addition, the very heavy top and bottom squarks of
focus point supersymmetry naturally produce a Higgs boson mass at or
above 115 GeV without requiring heavy gauginos.  We conclude with an
extended discussion of issues related to the definition of naturalness
and comment on several other prescriptions given in the literature.

\end{abstract}



\newpage

\renewcommand{\thefootnote}{\arabic{footnote}}
\setcounter{footnote}{0}

\section{Introduction}
\label{sec:introduction}

Among the motivations for supersymmetric extensions of the standard
model are three important virtues: they provide a natural solution to
the gauge hierarchy
problem~\cite{Maiani,Witten:1981nf,Veltman:1981mj,Kaul:1982wp}; they
predict a suitable particle candidate for cold dark
matter~\cite{Goldberg:1983nd,Ellis:1983wd}; and they incorporate the
unification of coupling constants~\cite{Dimopoulos:1981yj}.  All three
of these virtues are realized in a straightforward way if
superpartners masses are of order the weak scale.  At present,
however, no superpartners have been discovered at colliders. Even more
problematic, their virtual effects on low energy observables have also
not been seen.  The incompatibility of generic supersymmetric models
with low energy constraints encompasses a diverse set of difficulties,
which together are known as the supersymmetric flavor and CP problems.

Much of supersymmetric model building is motivated by the desire to
solve these problems without sacrificing some or all of the virtues
mentioned above.  There are many approaches to this puzzle.
Typically, the preservation of naturalness is assumed to require
superpartner masses below 1 TeV.  The supersymmetric flavor problems
are then solved, for example, by scalar degeneracy.  Dynamical
mechanisms guaranteeing scalar degeneracy have been found.  The
virtues of these mechanisms are many, but there are also typically a
number of attendant difficulties, such as the $\mu$ problem in
gauge-mediated models~\cite{Giudice:1999bp} and the problem of
tachyonic sleptons in anomaly-mediated models~\cite{Randall:1999uk}.
In addition, the suppression of CP violation usually requires
additional structure (see, for example,
Refs.~\cite{Randall:1999uk,Pomarol:1999ie,Katz:1999uw,Moroi:1999km,%
Feng:2000hg}), and the most natural dark matter particle, a neutralino
with the desired thermal relic density, is almost always eliminated
(although new dark matter candidates may
emerge~\cite{Dimopoulos:1996gy,Han:1997wn,Moroi:2000zb}).

Focus point supersymmetry~\cite{Feng:2000mn,Feng:2000zg,%
Feng:1999sw,Feng:2000hg,Agashe:2000ct} has been proposed as an
alternative to these approaches.  In focus point supersymmetry, all
squarks and sleptons, including those of the third generation,
naturally have masses well above 1 TeV.  All supersymmetric flavor and
CP problems are then ameliorated by decoupling, while preserving all
of the virtues listed above.  The naturalness of super-TeV scalars
arises from correlations among supersymmetry parameters.  More
specifically, the weak scale value of the parameter $m_{H_u}^2$, and
with it, the scale of electroweak symmetry breaking, is highly
insensitive to the values of the scalar masses $m_i$, and is
determined primarily by gaugino masses $M_i$ and trilinear scalar
couplings $A_i$.  Assuming a hierarchy $m_i \gg M_i, A_i$, as follows
naturally from, for example, an approximate
$R$-symmetry~\cite{Feng:1996dn}, the observed weak scale may then be
obtained without large fine-tuning, even in the presence of very large
scalar masses.

The conditions for the realization of focus point supersymmetry imply
testable correlations in the superpartner mass spectrum.  Sufficient
conditions have been presented in Ref.~\cite{Feng:2000mn}.  For
example, for {\em any} value of $\tb \agt 5$ and $\mt \approx
174~\gev$, a universal scalar mass guarantees focus point
supersymmetry.\footnote{In fact, focus point supersymmetry relies on
only a small subset of the universality assumption, being independent
of all scalar masses with small Yukawa
couplings~\cite{Feng:2000mn,Feng:2000zg}.  In addition, no relations
are required among the gaugino masses and $A$ parameters, and
supersymmetry breaking need not be gravity-mediated.  For these
reasons, focus point supersymmetry encompasses a broad class of
models, and may be found in models with gauge- and anomaly-mediated
supersymmetry breaking~\cite{Feng:2000hg,Agashe:2000ct}.}  The
simplicity of the required scalar mass boundary condition, and the
strong dependence of this simplicity on concrete experimental facts,
in particular, the measured top quark mass, provide two of the more
striking motivations for the framework.

In this respect, the motivation for focus point supersymmetry shares
many features with a well-known precedent --- the argument for
supersymmetric grand unified theories (GUTs).  Recall that, assuming
minimal supersymmetric field content, the renormalization group (RG)
trajectories~\cite{Georgi:1974yf} of the three standard model gauge
couplings focus to a point at the scale $\mgut \simeq 2 \times
10^{16}~\gev$~\cite{gaugeunif}.  This intersection is highly
non-trivial.  Assuming supersymmetric thresholds around the TeV scale,
there are no free parameters, and the meeting requires the standard
model gauge couplings to be within a few percent of their precisely
measured values.  This may be regarded as a coincidence.  However, it
may also be taken as evidence for supersymmetry with grand
unification, especially as grand unification provides a simple and
elegant explanation of the standard model gauge structure and
representation content~\cite{Georgi:1974sy}.  Indeed, this precise
quantitative success has been taken by some as an important advantage
of supersymmetry over all other attempts to address the gauge
hierarchy problem.

Focus point supersymmetry is motivated by a similar argument.
Assuming a universal GUT scale scalar mass, the family of $m_{H_u}^2$
RG trajectories for different values of this universal mass meet at a
point, the weak scale.  This meeting is also highly non-trivial.
Assuming unification at the GUT scale, there are no free parameters,
and the meeting requires the precisely measured top quark mass to be
within $\sim 2\%$ ($\sim 1\sigma$) of its measured
value~\cite{Feng:2000zg}.  This may be regarded as a coincidence.
However, it may also be taken as evidence for supersymmetry with a
large universal scalar mass, especially if it provides a simple
solution to the longstanding supersymmetric flavor and CP
problems.\footnote{This analogy highlights the similarity of focus
point supersymmetry and gauge coupling unification in their strong
dependence on precisely measured experimental data. Note, however,
that in the case of focus point supersymmetry, the meeting is of a
family of RG trajectories, of which only one can be realized in
Nature.}

In two studies with Wilczek~\cite{Feng:2000gh,Feng:2000zu}, we
explored the cosmological and astrophysical implications of focus
point supersymmetry.  In particular, we found that the focus point
framework preserves the most natural supersymmetric dark matter
candidate, the stable neutralino with the desired thermal relic
density.  However, unlike traditional scenarios in which this
neutralino is Bino-like, in focus point models it is a
gaugino-Higgsino mixture.  The Higgsino component has important
implications for dark matter searches.  For example, many indirect
detection signal rates are enhanced by several orders of magnitude.
The focus point scenario therefore predicts observable signals in
diverse experiments, ranging from neutrino and gamma-ray telescopes to
space-based searches for anti-particles in cosmic rays.

In this study, we address the following question: to what extent can
all of the supersymmetric flavor and CP problems be solved in focus
point supersymmetry by heavy scalars?  We consider the example of a
universal scalar mass in minimal supergravity.  In this simple
realization of focus point supersymmetry, many supersymmetric flavor
problems are solved by assumption.  However, even theories with
universal scalar masses generically violate current bounds on proton
decay and electric dipole moments, and they may also be significantly
constrained by measurements of the muon's magnetic dipole moment and
$\bsg$.  We will evaluate the status of focus point supersymmetry with
respect to each of these constraints in
Secs.~\ref{sec:proton}--\ref{sec:bsgamma}.

Of course, focus point supersymmetry also has important implications
for high energy colliders.  The prospects for discovering multi-TeV
squarks at the LHC have been considered in
Refs.~\cite{Allanach:2000ii,Chattopadhyay:2000qa}.  In
Sec.~\ref{sec:higgs} we consider the implications of focus point
supersymmetry for discovery of the light Higgs boson.  Of all of the
as-yet-undiscovered particles of the minimal supersymmetric model, the
Higgs boson is of special interest, given current stringent
constraints on its mass, the recently reported evidence for its
observation at LEP, and the prospect for discovery at Tevatron Run II.
Focus point supersymmetry differs from all other proposed
supersymmetric models in that {\em all} squarks and sleptons,
including the top and bottom squarks, may be naturally heavy.  For
this reason, focus point supersymmetry has novel implications for the
Higgs boson mass.  We will show that large radiative corrections from
super-TeV squarks naturally lead to Higgs masses in the experimentally
preferred range.

Finally, we close with an extended discussion of naturalness in
Sec.~\ref{sec:naturalness}.  While no discussion of naturalness and
fine-tuning can claim quantitative rigor, the possibility of focus
point supersymmetry raises a number of qualitatively novel issues.
When confronted with these issues, various naturalness prescriptions
in the literature yield qualitatively different results that should
not be dismissed as merely subjective ambiguities.  In this section,
we compare our approach with others currently in the literature to
clarify and highlight the essential differences.  We also reiterate
that the focus point is valid for all values of $\tb \agt 5$. In
Ref.~\cite{Feng:2000zg}, we demonstrated this analytically for
moderate $\tb$ and $\tb \approx m_t/m_b$ and numerically for all
$\tb$.  In the Appendix we supply the analytical proof for all $\tb
\agt 5$.

\section{Proton Decay and Gauge Coupling Unification}
\label{sec:proton}

Constraints from proton decay and the status of gauge coupling
unification are intimately connected.  As reviewed above, the apparent
unification of gauge couplings in supersymmetry has long been
considered an important virtue.  The advantage of the minimal
supersymmetric standard model over the standard model with respect to
gauge coupling unification is twofold.  First, the gauge couplings
unify more accurately.  This simplifies attempts to build GUT models,
since abnormally large threshold corrections are not required. Second,
the unification scale $\mgut$ is high enough that proton decay,
mediated by GUT scale particles~\cite{Dimopoulos:1982dw},
is sufficiently suppressed to evade experimental bounds.

The current status of supersymmetric unification is, however,
significantly more complicated.  Analyses of gauge coupling
unification now include two-loop RG equations~\cite{twoloop} and
leading-log~\cite{LL} and finite~\cite{Bagger:1995bw,finite} weak
scale threshold corrections.  In addition, the measurement of $\sin^2
\theta_W$ has improved.  With these refinements, the gauge couplings
are found to miss each other with a significant discrepancy.
Defining, as usual, the GUT scale through the relation $g_1(\mgut)
\equiv g_2(\mgut)$, a quantitative measure of the mismatch is
\begin{equation}
\varepsilon \equiv {g_3(\mgut)-g_1(\mgut)\over g_1(\mgut)} \ .
\label{epsilon}
\end{equation}
The parameter $\varepsilon$ depends only on measured standard model
quantities and the weak scale supersymmetric particle spectrum.

\begin{figure}[tbp]
\postscript{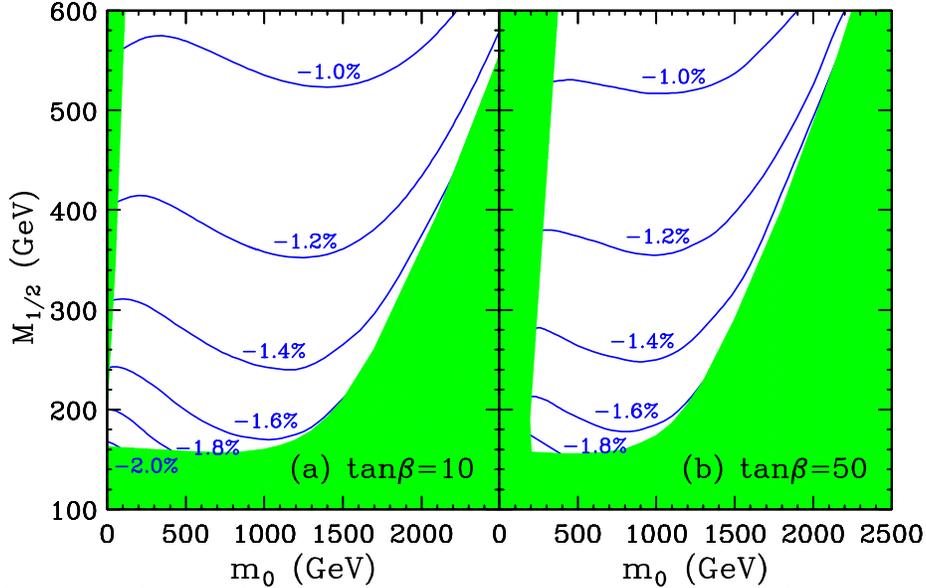}{0.74}
\caption
{Contours of $\varepsilon$, the GUT scale mismatch in gauge coupling
unification.  The shaded regions are excluded by the requirements of a
neutral LSP (left) and the 103 GeV chargino mass bound (right and
bottom). In this and all following plots unless otherwise noted, we
fix $\mt = 174~\gev$, $A_0=0$ and $\mu>0$, and choose representative
values of $\tb$ as indicated. }
\label{epsg}
\end{figure}

In minimal supergravity, the weak scale spectrum is fixed by 4+1
parameters: $m_0$, $\mgaugino$, $A_0$, $\tb$, and $\sign(\mu)$.  We
determine the weak scale theory by two-loop RG evolution with full
1-loop threshold corrections.  The magnitude of $\mu$ is determined by
(full one-loop) radiative electroweak symmetry breaking.  Values of
$\varepsilon$ in this framework are presented in Fig.~\ref{epsg}.
Heavy superpartners reduce
$|\varepsilon|$~\cite{LL,Bagger:1995bw,finite}.  One might therefore
naively expect $|\varepsilon|$ to be minimized in the focus point
region with large $m_0$.  For fixed $\mgaugino$, $|\varepsilon|$
indeed decreases as $m_0$ increases up to about 1 TeV. Above 1 TeV,
however, $|\mu|$ eventually drops, and threshold corrections from
light Higgsinos cause $|\varepsilon|$ to increase again.  As a result,
throughout parameter space, $-2\% < \varepsilon <
-1\%$~\cite{Bagger:1995bw,finite,Pierce:1997zz}.  This GUT scale
discrepancy is related to $\alpha_s(M_Z)$ by the approximate relation
\begin{equation}
\delta\alpha_s \approx 2 \frac{\alpha_s^2}{\alpha_G} \varepsilon
\approx 0.7\, \varepsilon \ .
\end{equation}
The current value of the strong coupling constant is $\alpha_s(M_Z) =
0.119 \pm 0.002$~\cite{Groom:2000in}.  In terms of the experimental
uncertainty, then, the mismatch is a $3.5\sigma$ to $7\sigma$ effect.

Of course, one might hope that the mismatch in couplings is a
reflection of GUT scale threshold corrections.  In the simple case of
minimal SU(5)~\cite{minimalsu5}, for example, the combined threshold
correction due to the colored Higgs bosons $H_3$, GUT scale gauge
bosons, and the Higgs bosons in the $\bold{24}$ representation
is~\cite{Hisano:1992mh,Hagiwara:1993ys,Yamada:1993kv}
\begin{equation}
\varepsilon_{H_3}= 0.3 \frac{\alpha_G}{\pi}
\ln\left( {M_{H_3}\over \mgut} \right) \ .
\label{H3threshold}
\end{equation}
We see that light colored Higgs bosons may explain the mismatch.
However, recent progress in proton decay, both experimental and
theoretical, places stringent lower limits on GUT scale particle
masses.  Recent results from Superkamiokande significantly strengthen
limits on the proton lifetime.  In the $p\to K^+ \bar{\nu}$ channel,
for example, the current limit is $\tau(p \to K^+ \bar{\nu}) > 1.9
\times 10^{33}~\text{yr}$~\cite{SuperK}.  On the theoretical side, it
is now known that there are dimension 5 supersymmetric contributions
to proton decay involving right-handed scalars (the so-called RRRR
operators) with amplitudes that scale as
$\tan^2\beta$~\cite{Lucas:1997bc,Goto:1999qg,Babu:1998ep,Goto:1999iz}.
The combined effect of these developments is that the colored Higgs
mass $M_{H_3}$ is typically required to be far above $\mgut$,
especially for large $\tb$, and the mismatch in gauge couplings is, in
fact, exacerbated by such GUT scale threshold corrections.

In non-minimal models there will be additional GUT threshold
corrections.  These corrections may improve the unification of
couplings~\cite{missingpartner,Lucas:1996ic,Barr:1999dw}, make it even
more problematic~\cite{Hall:1995eq,Barr:1997hq}, or be sufficiently
complicated that no definite statement can be made~\cite{anysize}. In
general, we may write the total GUT threshold correction
as~\cite{Lucas:1996ic,Barr:1999dw}
\begin{equation}
\varepsilon_{H_{\text{eff}}} + \Delta \varepsilon 
\equiv 0.3 \frac{\alpha_G}{\pi}
\ln\left( {M_{H_{\text{eff}}}\over \mgut} \right)
+ \Delta \varepsilon \ ,
\label{Deltaepsilon}
\end{equation}
where $M_{H_{\text{eff}}}$ is the {\em effective} color triplet Higgs
mass entering the proton decay amplitude, and $\Delta \varepsilon$ is
the threshold correction from sectors of the theory that have no
impact on proton decay.  $\Delta \varepsilon$ is generically a
model-dependent holomorphic function of ratios of GUT scale masses and
vacuum expectation values. For some models, however, $\Delta
\varepsilon$ simplifies tremendously.  For example, in missing partner
SU(5) models~\cite{missingpartner}, $\Delta \varepsilon = 0.3
[\alpha_G/\pi] [15 \ln 2 - (25/2) \ln 5 ] \sim
-3.9\%$~\cite{Hagiwara:1993ys,Yamada:1993kv}, and in the complete
SO(10) model of Ref.~\cite{Hall:1995eq}, $\Delta \varepsilon = 0.3
[\alpha_G/\pi]\, 21 \ln 2 \sim +5.8\%$~\cite{Lucas:1996ic}.  In both
cases, there are no remaining free parameters.

These examples illustrate that the severity of the proton decay
problem is model dependent; some specific models may even be
consistent with current constraints. However, it is clear, as has
recently been emphasized in Ref.~\cite{Dermisek:2000hr}, that
generally speaking, current proton decay bounds place a significant
strain on many well-motivated models, as they exclude the large
threshold corrections necessary for gauge coupling unification.
General mechanisms for suppressing proton decay are therefore welcome,
in that they allow greater freedom in GUT model building.

\begin{figure}[tbp]
\postscript{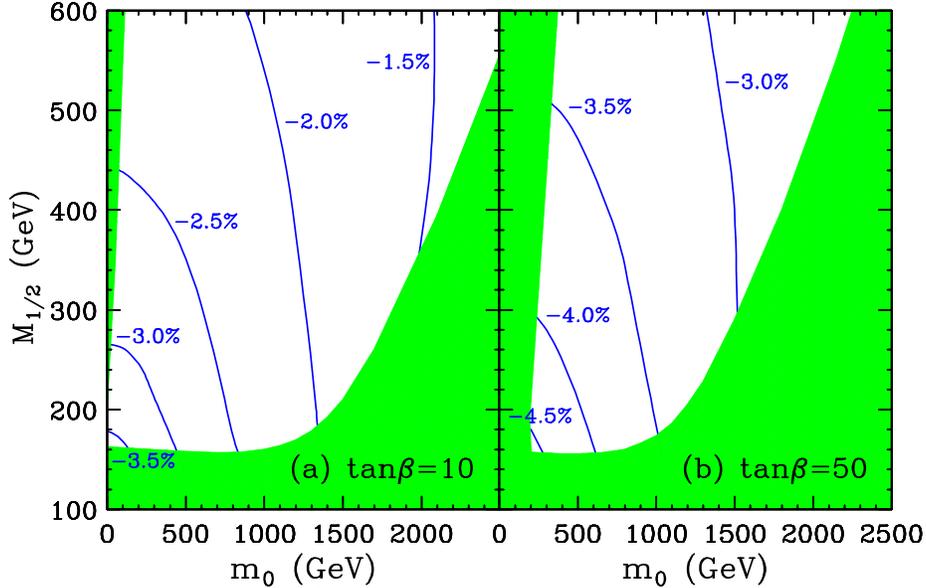}{0.74}
\caption{Contours of the minimal (in absolute value) threshold
correction $\Delta \varepsilon$ from non-minimal GUT particle content
allowed by coupling constant unification and current proton decay
limits.}
\label{fig:deltaeps}
\end{figure}

In this spirit, we now investigate the implications of focus point
supersymmetry.  In Fig.~\ref{fig:deltaeps}, we plot
\begin{equation}
\Delta \varepsilon \equiv \varepsilon - \varepsilon_{H_{\text{eff}}} 
\ ,
\end{equation}
where $\varepsilon$ is defined in Eq.~(\ref{epsilon}), using weak
scale experimental inputs and sparticle spectra, and
$M_{H_{\text{eff}}}$ is taken to be as low as possible consistent with
current proton lifetime bounds.\footnote{In evaluating the proton
lifetime bound, we use the fits of Ref.~\cite{Goto:1999iz}, which
include the RRRR contributions.} In other words,
Fig.~\ref{fig:deltaeps} shows the minimal (in absolute value)
threshold correction from non-minimal GUT particle sectors allowed by
coupling constant unification and current proton decay constraints.

We see that in the focus point region with large $m_0$, coupling
constant unification may be achieved with smaller non-minimal
threshold corrections. In this region, proton decay is highly
suppressed by heavy squarks and sleptons.  The allowed value of
$M_{H_{\text{eff}}}$ is therefore lower than in conventional models,
and the required additional GUT threshold correction from non-minimal
GUT sectors is reduced.  More quantitatively, for a fixed $\mgaugino$,
the required threshold corrections are decreased by 1\% to 1.5\% for
focus point scenarios with multi-TeV scalars relative to conventional
scenarios with $m_0 \sim {\cal O}(100 ~\gev)$. Thus in many GUTs, the
prediction for $\alpha_s(M_Z)$ is closer to the experimental value by
3$\sigma$ to 5$\sigma$ in focus point models relative to conventional
scenarios.  As a result, in focus point scenarios, large threshold
corrections from baroque non-minimal sectors are not required,
increasing the viability of simpler and, presumably, more credible
models.

\section{Electric Dipole Moments}
\label{sec:EDMs}

Constraints on CP violation can be flavor-violating, as in the case of
$\epsilon_K$, or flavor-conserving, as in the case of electric dipole
moments (EDMs). For generic theories, the bound from $\epsilon_K$ is
the most stringent of all flavor- and CP-violating constraints.
However, the $\epsilon_K$ constraint is satisfied in many theories
with natural flavor violation suppression.  In contrast, the EDM
constraints are more robust, in the sense that they cannot be avoided
simply by scalar degeneracy or alignment.  For this reason, EDMs pose
a serious problem even in models with a universal scalar mass, as well
as in gauge- and anomaly-mediated theories. EDMs have been studied in
many supersymmetric models. (See, for example,
Ref.~\cite{Moroi:1999km} and references therein.)  Here we evaluate
the predictions for EDMs in focus point supersymmetry.

In minimal supergravity, the parameters $\mgaugino$, $A_0$, $\mu$, and
$B$ may all be complex.  The first two are input parameters of the
framework.  The $\mu$ and $B$ parameters are constrained by
electroweak symmetry breaking, but this restricts only their
magnitudes.  In principle, it is possible that the phases of these
four parameters are related, but lacking any specific mechanism for
their generation, we treat them as independent.  The freedom of
U(1)$_R$ and U(1)$_{PQ}$ rotations imply that only two phases are
physical.  One of these is
\begin{equation} 
\thetacp \equiv {\rm Arg}(\mu B^* \mgaugino) \ ,
\end{equation}
which generates EDMs.  The EDM $d_f$ of fermion $f$ is the coefficient
of the electric dipole term
\begin{equation} 
{\cal L}_{\rm EDM} = -\frac{i}{2} d_f \ \bar{f}
\sigma^{\alpha \beta} \gamma_5 f \ F_{\alpha \beta} \ ,
\label{edm}
\end{equation} 
where $F$ is the electromagnetic field strength.  Supersymmetric
contributions to EDMs arise from sfermion-gaugino loops.  As is clear
from the structure of the operator in Eq.~(\ref{edm}), these
contributions require a chirality flip along the fermion-sfermion
line.  For down-type fermions, these contributions are therefore
enhanced for large $\tb$.

We will consider the stringent constraints from the EDMs of the
electron and neutron. (The EDM of the mercury atom is also competitive
in some regions of parameter space~\cite{Falk:1999tm}.)  For the
electron, there is a direct $\tb$ enhancement.  This is most easily
seen in the mass insertion approximation, where, for large $\tb$, the
supersymmetric contributions take the form
\begin{equation}
  d_e^{\rm SUSY} \approx 
\sin\thetacp \frac{m_e}{2} \mu \tb \left[ 
g_1^2 M_1 F_1 (M_1^2, \mu^2, m_{\tilde{e}_L}^2, m_{\tilde{e}_R}^2)
+ g_2^2 M_2 F_2 (M_2^2, \mu^2, m_{\tilde{e}_L}^2, m_{\tilde{\nu}_e}^2)
\right] \ ,
\end{equation}
where explicit formulae for the $F$ functions are given in
Ref.~\cite{Moroi:1996yh}.  For large sfermion masses, $F \sim
m_{\tilde{f}}^{-4}$.  To calculate the neutron EDM, we must model the
structure of the neutron.  We adopt the non-relativistic quark model,
in which the neutron EDM is $d_n = (4d_d - d_u)/3$.  Contributions to
the quark EDMs are similar to those for the electron, with the
exception that there are additional contributions from squark-gluino
diagrams.  Note that, since $d_d \propto \tb$, the neutron EDM is also
enhanced for large $\tb$.

The standard model predicts vanishing EDMs, to foreseeable
experimental accuracy.  At present, no anomaly is seen in EDM
measurements.  {}From the measurement $d_e=(0.18 \pm 0.12 \pm 0.10)
\times 10^{-26}e~\cm$~\cite{Commins:1994gv}, we obtain the constraint
$|d_e| \leq 0.44 \times 10^{-26}\ e~\cm$, where the right-hand side is
the upper bound on $|d_e|$ at 90\%~C.L.  For the neutron, the current
90\%~C.L. limit is $|d_n| \leq 0.63 \times 10^{-25}\
e~\cm$~\cite{Harris:1999jx}.

\begin{figure}[tbp]
\postscript{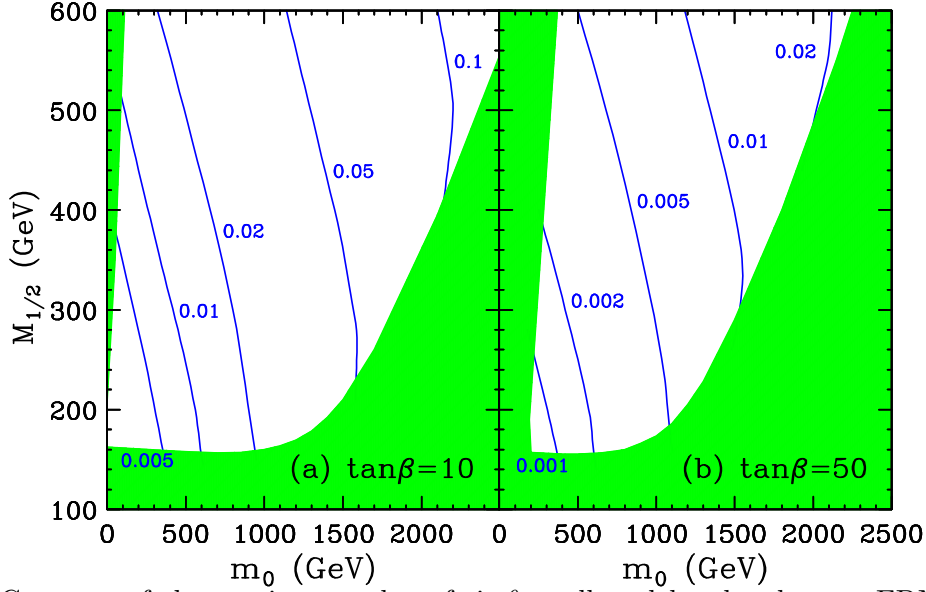}{0.74}
\caption
{Contours of the maximum value of $\sin\thetacp$ allowed by the
electron EDM constraint $|d_e|\le 0.44\times 10^{-26}\ e~\cm$.}
\label{edm_de}
\end{figure}

\begin{figure}[!tb]
\postscript{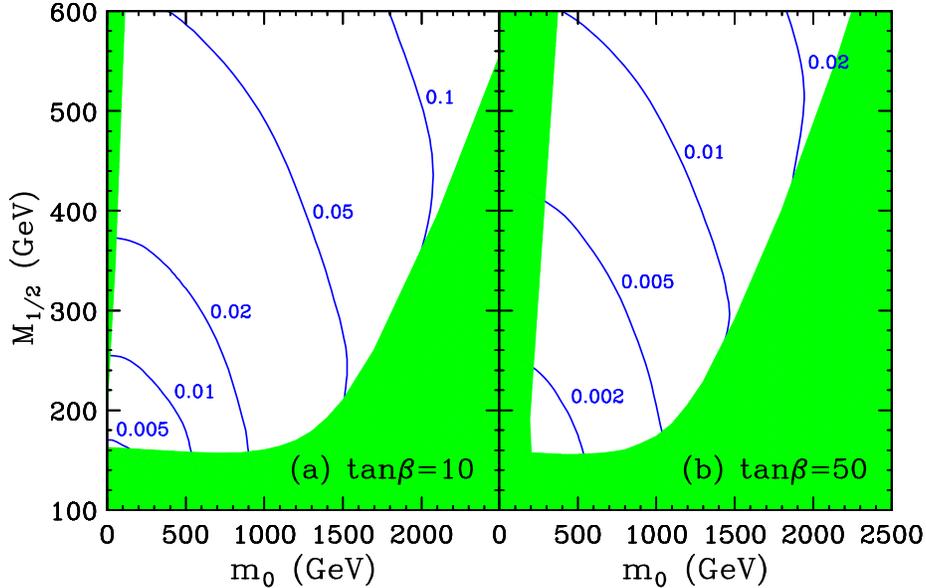}{0.74}
\caption
{As in Fig.~\ref{edm_de}, but for the neutron EDM constraint $|d_n|\le
0.63\times 10^{-25}\ e~\cm$.}
\label{edm_dn}
\end{figure}

The present constraints on EDMs severely restrict the possible values
of $\thetacp$. In Figs.~\ref{edm_de} and \ref{edm_dn}, we plot the
maximal allowed values of $\thetacp$ given the constraints of the
electron and neutron EDMs, respectively.  The EDMs are calculated in
the exact mass eigenstate basis. We see that current constraints from
the electron and neutron are roughly comparable.  For sub-TeV values
of $m_0$, $\thetacp$ is constrained to be less than of order $10^{-3}$
to $10^{-2}$, depending on $\tb$.  In the absence of an understanding
of the origin of this phase, this appears to require a strong
fine-tuning.  For the focus point scenario with multi-TeV $m_0$, these
constraints may be relaxed by over an order of magnitude.  In the $\tb
= 10$ case, ${\cal O}(0.1)$ phases are allowed.

\begin{figure}[tbp]
\postscript{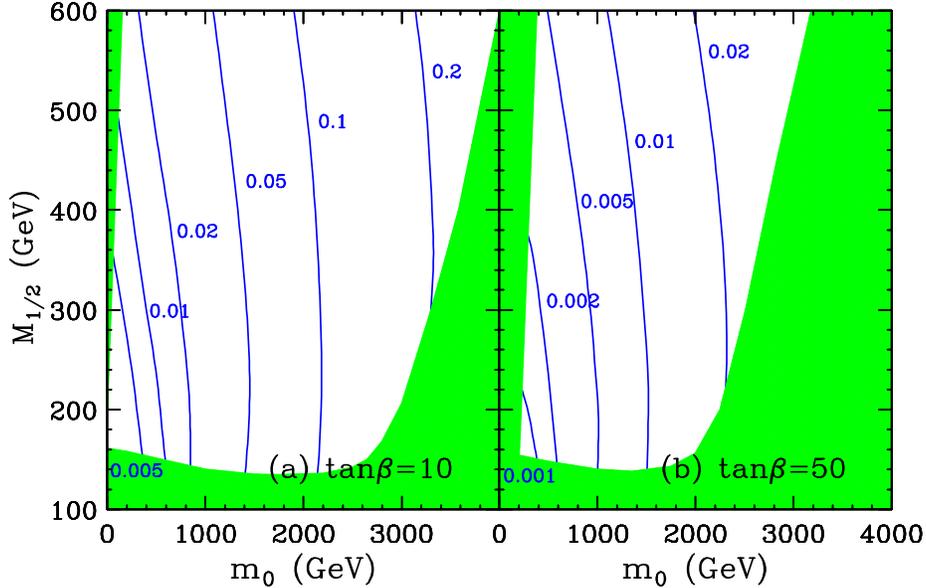}{0.74}
\caption
{As in Fig.~\ref{edm_de}, but for $m_t = 179~\gev$.}
\label{edm_de_179}
\end{figure}

It is important to note that there is some sensitivity to the assumed
top mass.  For larger top quark mass, but still within the
experimental bounds, the excluded region from chargino mass limits
moves to larger $m_0$, and so even larger scalar masses are allowed.
In Fig.~\ref{edm_de_179}, we show the $CP$-violating phases allowed by
the electron EDM, but with an assumed top quark mass of $\mt =
179~\gev$, within the 1$\sigma$ experimental bound.  As $m_0$ now
extends to over 3 TeV, even larger phases are allowed.  A similar
improvement is found in the neutron EDM case.

\section{Magnetic Dipole Moment of the Muon}
\label{sec:MDM}

Supersymmetric particles also contribute radiatively to magnetic
dipole moments (MDMs). Such contributions are even more robust than
EDMs, as they require neither CP nor flavor violation.  At present the
most stringent constraint comes from the muon's anomalous MDM $a_\mu =
\frac{1}{2} (g-2)_\mu$, which is the coefficient of the operator
\begin{equation}
  {\cal L}_{\rm MDM} = a_\mu \frac{e}{4m_\mu} \
  \bar{\mu} \sigma^{\alpha\beta} \mu \ F_{\alpha\beta} \ .
\end{equation}

The supersymmetric contributions are similar to those discussed above
for the electron EDM, arising from slepton-neutralino and
sneutrino-chargino loops.  In the large $\tb$ regime, they take the
form
\begin{equation}
a_\mu^{\text{SUSY}} \approx 
m_\mu^2 \ \mu \tb \left[ 
g_1^2 M_1 F_1(M_1^2,\mu^2,m_{\tilde{\mu}_L}^2,m_{\tilde{\mu}_R}^2)
+ g_2^2 M_2 F_2(M_2^2,\mu^2,m_{\tilde{\mu}_L}^2,m_{\tilde{\nu}_\mu}^2)
\right] \ ,
\end{equation}
where in this section, we assume all parameters real.  The $F$
functions are as in Eq.~(\ref{edm}).

The anomalous MDM of the muon has been measured at
CERN~\cite{Bailey:1979mn} and Brookhaven~\cite{Carey:1999dd,Carey}.
The current world average is $a_\mu^{\rm exp} = (116\ 592\ 05\pm
45)\times 10^{-10}$~\cite{Carey}, consistent with the standard model.
The uncertainty is statistics dominated, and will be reduced by the
ongoing Brookhaven experiment E821.  With data already being
collected, the uncertainty should be reduced to $\sim 7 \times
10^{-10}$, and the ultimate goal of E821 is $\Delta a_{\mu} \sim 4
\times 10^{-10}$~\cite{GrossePerdekamp:1999up}.  The current standard
model prediction for the muon's anomalous MDM is $a_\mu^{\rm th} =
(116\ 591\ 62\pm 8)\times 10^{-10}$~\cite{Hughes:1999fp}.  The
uncertainty in the prediction is dominated by the difficulty of
evaluating the hadronic vacuum polarization contribution, but is being
reduced by improved low energy data.  If the theoretical prediction is
brought under control, a reasonable $2\sigma$ limit in the near future
is $8\times 10^{-10}$.

\begin{figure}[tbp]
\postscript{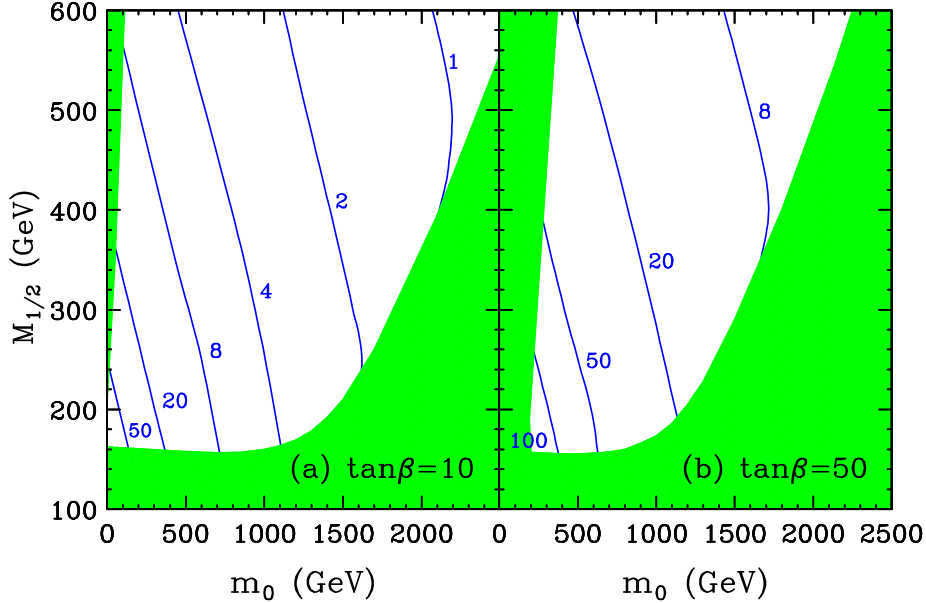}{0.74}
\caption{The muon anomalous MDM $a_\mu^{\rm SUSY}$ in units of
$10^{-10}$. }
\label{fig:mdm}
\end{figure}

The supersymmetric contribution to the muon anomalous MDM $a_\mu^{\rm
SUSY}$, in the mass insertion approximation~\cite{Moroi:1996yh}, is
given in Fig.~\ref{fig:mdm}.  As expected, the contribution is
enhanced for large $\tb$, and highly suppressed by heavy sleptons in
the focus point region.  A measured deviation is consistent with focus
point supersymmetry, but only for large $\tb$.  On the other hand, if
no deviation is found, the muon's anomalous MDM will be a strong
argument for heavy superpartners.  (Recall that the muon MDM is
flavor- and CP-conserving, so cannot be eliminated by, for example,
scalar degeneracy or small phases.)  For moderate $\tb$,
considerations of dark matter relic density eliminate the moderate
$m_0$ possibility (see below), and so a muon MDM consistent with the
standard model would require $m_0$ above a TeV.  For large $\tb$, even
such a robust cosmological constraint is unnecessary: for $\tb=50$,
the absence of an anomaly would require $m_0 \agt 1.5~\tev$, well into
the focus point region.

\section{\bold{\bsg}}
\label{sec:bsgamma}

It is well-known that the supersymmetric contributions to $\bsg$ may be
large.  In the standard model, this flavor-violating transition takes
place only at one-loop through a $W$ boson.  In supersymmetric
theories, there are a variety of additional one-loop
contributions~\cite{Bertolini:1991if}, most importantly those from
charged Higgs- and chargino-mediated processes.  These are both
enhanced by large $\tb$ in focus point supersymmetry.  For the
chargino diagrams, this is true for the standard reason of enhanced
Yukawa couplings.  For the charged Higgs diagram, it holds because
large $\tb$ implies small charged Higgs masses. At small $\tb$,
$m_{H^+}$ is of order the scalar superpartner masses, and so is well
above 1 TeV in the focus point region.
However, for large $\tb$, by the approximate up-down
symmetry, both $m_{H_u}^2$ and $m_{H_d}^2$ have weak scale focus
points, and so $m_{H^+}$ is typically of the order of 100 GeV.

We evaluate $B(B \to X_s \gamma)$ as follows.  As we are primarily
interested in the case where there is a hierarchy between the
scale of the superpartner masses $\msusy$
and the weak scale $\mweak$, we are careful {\em not} to
decouple the supersymmetric contributions at the weak scale,
as is usually done. Instead, we need to resum the large logarithms of
$\msusy/\mweak$ \cite{Anlauf:1994df,Bagger:1997gg,Degrassi:2000qf}.
Operationally, we evaluate the leading order
supersymmetric contributions~\cite{Bertolini:1991if} at the
superpartner scale (defined as the geometric mean of the two top
squark masses) and evolve them to the weak scale, using leading order
anomalous dimension coefficients.  At the weak scale, we match these
contributions to the effective Hamiltonian

\begin{equation}
{\cal H}_{\text{eff}} = -\frac{4G_F}{\sqrt{2}} V^*_{ts} V_{tb}
\sum_{i=1}^8 C_i {\cal O}_i \ .
\end{equation}
We use next-to-leading order matching conditions for the standard
model~\cite{NLOSM} and charged Higgs~\cite{NLOH+} contributions.

The weak scale Wilson parameters $C_i$ must then be evolved to the
low energy scale $\mu_b$ with the NLO anomalous dimension
matrix~\cite{Chetyrkin:1997vx}, and $B(B \to X_s \gamma)$ is then
evaluated using NLO matrix elements~\cite{bsgmatrix}, incorporating
the leading order QED and electroweak radiative
corrections~\cite{bsgQED,Kagan:1999ym}.  These results have been
included in a simple parameterization of Ref.~\cite{Kagan:1999ym},
which we adopt, taking $\mu_b = m_b$ and a photon energy cutoff
parameter $\delta = 0.9$.

The best current measurements of $B\rightarrow X_s\gamma$ from
CLEO~\cite{Ahmed:1999fh} and ALEPH~\cite{Barate:1998vz} may be
combined in a weighted average of $B(B\rightarrow X_s\gamma)_{\rm exp}
=(3.14\pm 0.48)\times 10^{-4}$~\cite{Kagan:1999ym}.  It is expected
that these measurements will be significantly improved at the $B$
factories, where large samples of $B$ mesons will greatly reduce
statistical errors.  At present, the standard model prediction is
$B(B\rightarrow X_s\gamma)_{\rm SM} =(3.29\pm 0.30)\times
10^{-4}$~\cite{Kagan:1999ym}.  The theoretical uncertainty is less
likely to improve substantially. We estimate that in the near future,
both theoretical and experimental uncertainties will be $\sim 0.3
\times 10^{-4}$. Combining these errors linearly, the resulting
$2\sigma$ limit will be $2.1\times 10^{-4} < B(B\rightarrow X_s\gamma)
< 4.5\times 10^{-4}$.

\begin{figure}[tbp]
\postscript{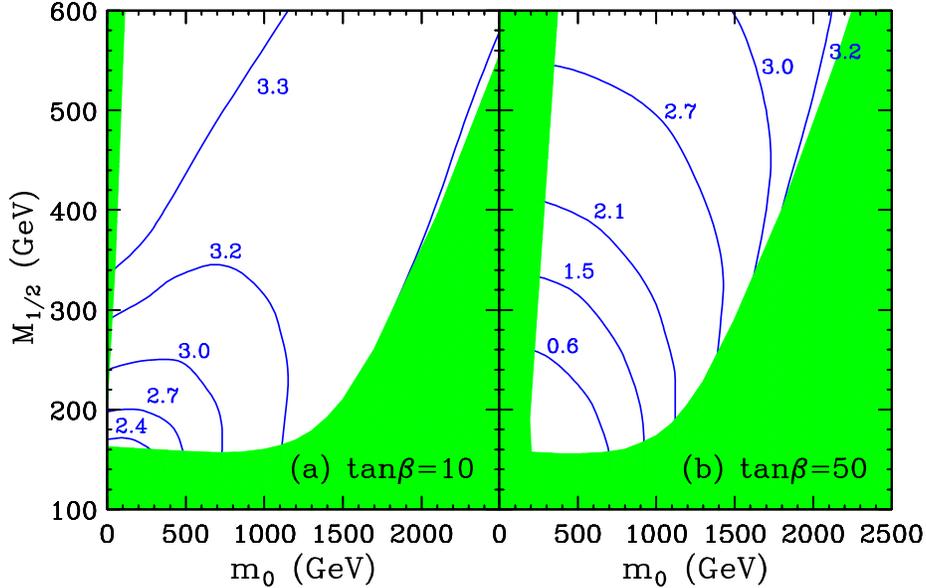}{0.74}
\caption{$B(B\rightarrow X_s\gamma)$ in units of $10^{-4}$ for $\mu >
0$.}
\label{fig:bsgp}
\end{figure}

\begin{figure}[htbp]
\postscript{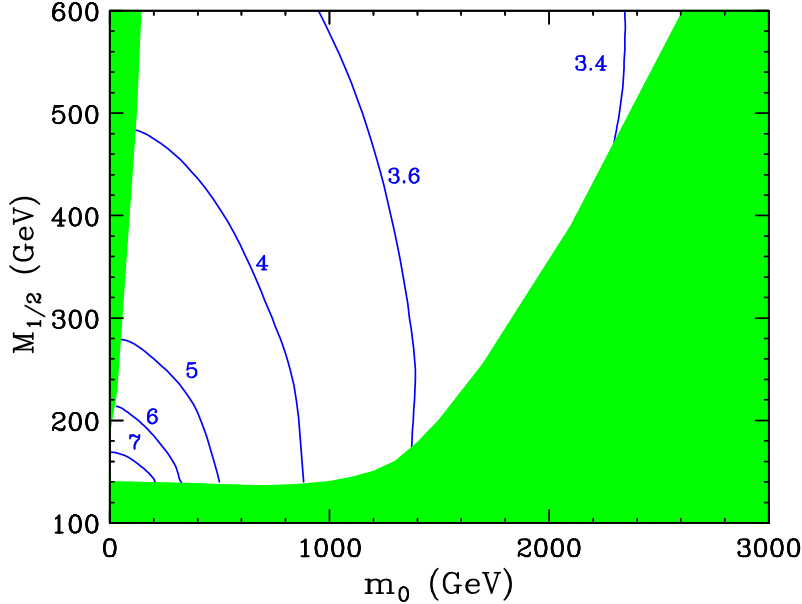}{0.64}
\caption{As in Fig.~\ref{fig:bsgp}, but for $\mu < 0$ and $\tb = 10$.}
\label{fig:bsgn}
\end{figure}

In Figs.~\ref{fig:bsgp} and \ref{fig:bsgn}, we plot contours of
$B(B\rightarrow X_s\gamma)$ for positive and negative $\mu$,
respectively.\footnote{We do not show results for $\tan\beta=50$ and
$\mu>0$. For such parameters, difficulties in obtaining correct
electroweak symmetry breaking exclude much of the parameter space, and
for the remaining region, the prediction for $\bsg$ is always very
large and excluded by current bounds.}  The charged Higgs contribution
is always constructive with the standard model. For $\mu < 0$, the
chargino contribution is also constructive, and predicted $\bsg$ rates
are enhanced.  For $\mu > 0$, the chargino contribution flips sign,
and may cancel the charged Higgs contribution.

We see that for the foreseeable future, focus point supersymmetry
predicts no measurable deviation from the standard model, and both
positive $\mu$ and negative $\mu$ (with moderate $\tb$) are consistent
if no deviation is seen. Of course, if no deviation is found,
supersymmetric models with sub-TeV scalar mass $m_0$ are also
consistent for moderate and low $\tb$.

\section{The Mass of the Light Higgs Boson}
\label{sec:higgs}

The implications of focus point supersymmetry for the light Higgs
boson are of special interest, given the present bound of $m_h >
113.5~\gev$, and the recent observation at LEP of a $2.9\sigma$ excess
of events consistent with the production of a standard model-like
Higgs boson with mass $m_h=115$ GeV~\cite{Igo-Kemenes,LEPHIGGS}.

As is well-known, in the minimal supersymmetric model, the light Higgs
boson mass satisfies $m_h \alt 130~\gev$.  This limit is saturated in
regions of parameter space, where, for example, trilinear $A$
parameters are adjusted to give maximal left-right scalar mixing in
the third generation squarks.  Such regions are, however,
extraordinarily unnatural, requiring extreme fine-tuning in the
electroweak potential~\cite{Feng:2000zg}.  In fact, in natural regions
of parameter space with $m_0 \alt 1$ TeV, Higgs boson masses as high
as those presently preferred are already highly constraining.  In
Ref.~\cite{Ellis:2000sv}, the authors concluded that a Higgs mass of
115 GeV, along with the assumption of a suitable Bino-like dark matter
candidate, implied lower limits on gaugino masses, with strong
(negative) implications for supersymmetry searches at the Tevatron.
In Ref.~\cite{Kane:2000kc}, similar considerations led the authors to
consider, among other possibilities, large CP violating phases, which
much necessarily cancel to high accuracy in EDMs.

\begin{figure}[tbp]
\postscript{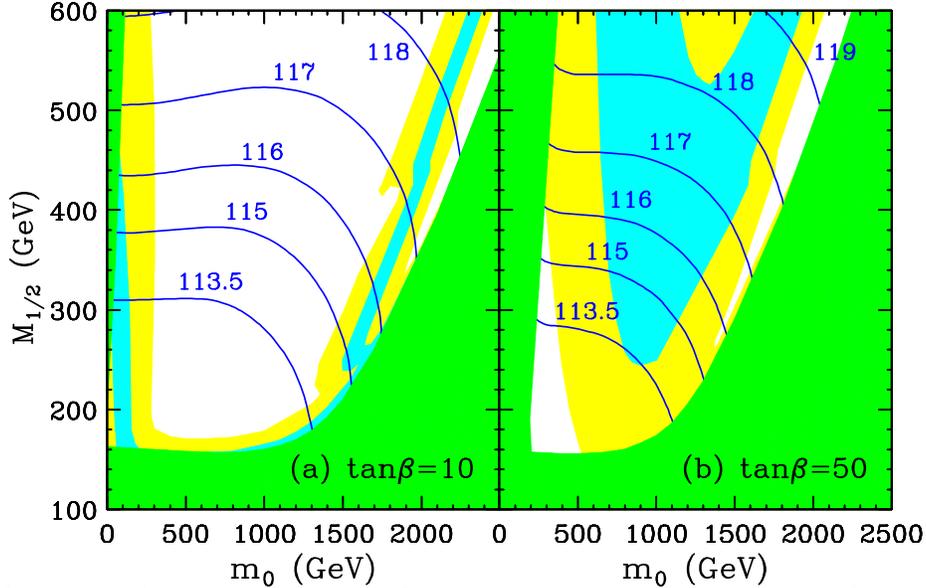}{0.74}
\caption
{Contours of Higgs mass $m_h$ in GeV.  Regions with allowed and
preferred dark matter relic density are also shown.  In the light
(yellow) shaded region, the thermal relic density of the neutralino
LSP is $0.025 \protect \alt \Omegachi h^2 \protect \alt 1$, and in the
dark (blue) shaded region it is in the preferred range $0.1 \protect
\alt \Omegachi h^2 \protect \alt 0.3$.  The unshaded region above the
preferred band has $\Omegachi h^2 \protect \agt 1$ and is excluded.}
\label{fig:mhiggs}
\end{figure}

\begin{figure}[htbp]
\postscript{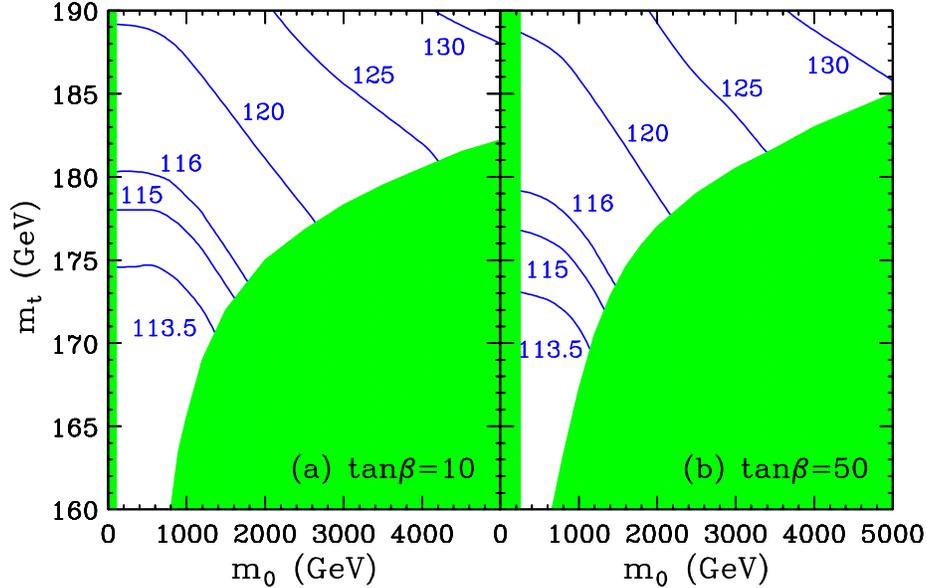}{0.74}
\caption{Contours of Higgs boson mass $m_h$ in GeV in the $(m_0, \mt)$
plane for fixed $\mgaugino=300~\gev$, $A_0 = 0$, and $\mu>0$.}
\label{fig:mh_m0mt}
\end{figure}

In focus point supersymmetry, {\em all} squarks and sleptons,
including those of the third generation, may be above 1 TeV without
significantly increased fine-tuning in the electroweak potential.
This is in contrast to all other proposed models, including those that
also make use of RG effects to resolve the tension between low energy
constraints and naturalness, but which, while allowing heavy first and
second generation scalars, require light third generation
superpartners~\cite{superheavy}.  As the dominant radiative
contributions to the light Higgs boson mass are logarithmically
dependent on top and bottom squark masses, this fact has strong
implications for the Higgs boson.  In Fig.~\ref{fig:mhiggs}, we
present contours of constant Higgs mass, including the full one-loop
radiative corrections as in Ref.~\cite{Pierce:1997zz}.  We see that
Higgs masses at or above 115 GeV are naturally and simply accommodated
in the focus point region.  In fact, a Higgs boson with mass
consistent with present bounds is an inescapable consequence of focus
point supersymmetry with multi-TeV squarks.  Varying $A_0$ within a
generous range allowed by naturalness does not change these
conclusions~\cite{Feng:2000zg}. In Fig.~\ref{fig:mh_m0mt} we
illustrate the dependence on the top quark mass.  Variations of $\mt$
within its 1$\sigma$ experimental uncertainty give rise to $\sim
2~\gev$ variations in $m_h$.

Note that the focus point region possesses a suitable neutralino dark
matter candidate, a Higgsino-gaugino mixture.  In
Fig.~\ref{fig:mhiggs}, we show also the regions with good thermal
relic density.  In conventional scenarios, with $m_0 \alt 1~\tev$
assumed, the LSP is Bino-like, and the thermal relic density
constrains $m_0$ to values of at most $\sim 200~\gev$.  The radiative
corrections from $m_0$ to the Higgs boson mass are therefore small,
and present bounds already require large $\mgaugino$.  However, as
noted in Refs.~\cite{Feng:2000gh,Feng:2000zu}, the assumption of a
Bino-like LSP is far from robust, and is violated even in the simple
framework of minimal supergravity.  {}From Fig.~\ref{fig:mhiggs}, we
see that a cosmologically attractive region exists in the focus point
region, with $m_0 > 1~\tev$.  In this region, the LSP is a
gaugino-Higgsino mixture, and its relic density may also be in the
preferred range $0.1 \alt \Omegachi h^2 \alt 0.3$.  The focus point
region therefore provides an excellent dark matter candidate in which
the Higgs boson mass is naturally in the currently preferred range.

\section{On Naturalness}
\label{sec:naturalness}

In the preceding sections, we have found several phenomenological
virtues of focus point scenarios with respect to proton decay, the
supersymmetric flavor and CP problems, and light Higgs boson
mass. Such attractive features would be offset by the ugliness of a
fine-tuned electroweak scale, were it not for the focus point
mechanism, which makes heavy scalars natural.  In this section, we
attempt to clarify several issues concerning naturalness by comparing
our prescription with several others in the literature.  Naturalness
has been discussed in a large number of studies.  In the following, we
do not attempt a comprehensive review, but rather highlight various
similarities and differences between our prescription and selected
other studies~\cite{Ellis:1986yg,Barbieri:1988fn,AC,RS,%
Romanino:1999ut,Ross:1992tz,deCarlos:1993yy,Chankowski}.

All definitions of naturalness are open to quantitative ambiguities.
However, this fact should not be allowed to obscure the many strong
qualitative differences that, as we will see, exist between various
naturalness prescriptions.  The claim that the focus point renders
multi-TeV scalars natural is {\em qualitatively} novel, and leads to
qualitatively new implications for many searches for supersymmetry.
For this reason, it is worthwhile to identify and explore the
underlying differences between our prescription and others in the
literature. As a by-product, we also highlight many issues in defining
naturalness that are seldom addressed.

\subsection{Our Prescription}
\label{sec:our}

We begin by briefly reviewing our naturalness prescription.  Readers
interested in a more careful and detailed description are referred to
Refs.~\cite{Feng:2000mn,Feng:2000zg}.  The five step prescription is
the following:

\noindent (1) Choose a supersymmetric model framework.  For example,
if one chooses minimal supergravity, one assumes input parameters
$\{m_0, \mgaugino, A_0, \tb, \sign(\mu)\}$ and adopts all the
assumptions encapsulated in these 4+1 parameters.

\noindent (2) For a given set of input parameters, determine all weak
scale parameters of the theory consistent with experimental data and
RG evolution.

\noindent (3) Choose some set of parameters to be free, continuously
variable, independent, and fundamental. In minimal supergravity, we
choose the GUT scale parameters $\{ a_i \} = \{ m_0, \mgaugino, A_0,
B_0, \mu_0\}$.  Note that we have included all parameters expected to
be intimately related to supersymmetry breaking, but none of the
others.

\noindent (4) For each fundamental parameter, define the sensitivity
coefficient\footnote{This formula corrects a typographical error in
the definition of $c_i$ in
Refs.~\cite{Feng:2000mn,Feng:2000zg}.}~\cite{Ellis:1986yg}
\begin{equation}
c_i \equiv 
\left| \frac{\partial \ln m_Z}{\partial \ln a_i} \right| 
= \left| \frac{a_i}{m_Z} \frac{\partial m_Z}{\partial a_i} \right| 
\ . 
\label{sensitivity}
\end{equation}

\noindent (5) Finally, define the overall measure of fine-tuning to be
\begin{equation}
c = \max \{ c_i \}\ .
\label{finetune}
\end{equation}

\subsection{Sensitivity Coefficients}

The sensitivity coefficients of Eq.~(\ref{sensitivity}) are the kernel
of most naturalness prescriptions.  They were first advanced as a tool
for quantifying naturalness by Ellis, Enqvist, Nanopoulos, and
Zwirner~\cite{Ellis:1986yg}.  These authors analyzed an $E_6$ model
with superpotential $W=h_t Q_3 U_3^c H + \lambda H \bar{H} N + k D
D^c$, where the first term is the top quark Yukawa coupling, $N$ is a
singlet Higgs field, $H$ and $\bar{H}$ are the standard Higgs
doublets, and $D$ and $D^c$ are exotic down-type quarks.  They then
defined sensitivity coefficients $c_i = \left| \partial \ln x /
\partial \ln a_i \right|$, where $x \equiv \langle N \rangle / \langle
H \rangle$, and used $c_\lambda, c_k < 5$ as a reasonable requirement
for natural regions of parameter space.

Aside from a difference in the framework being examined, our
naturalness definition differs from this one only in what parameter
has been chosen to represent the weak scale (their parameter $x$
vs. our $m_Z$).  This difference is minimal, and these prescriptions
are identical in spirit. Note that the sensitivity to the standard
model parameter $h_t$ was not included.

\subsection{Model Dependence and the Choice of Fundamental Parameters}

In another pioneering study, the sensitivity coefficients were then
used by Barbieri and Giudice to examine naturalness in the context of
minimal supergravity~\cite{Barbieri:1988fn}.  In that paper, an
overall fine-tuning parameter similar to that defined in
Eq.~(\ref{finetune}) was used: the sensitivities to all
supersymmetry-breaking parameters (and $\mu_0$) were included, but
sensitivities to standard model parameters were not.  These authors
considered a range of $h_t$ and ignored the effects of $h_b$.  For
particular $h_t$, the weak scale was found to be insensitive to
variations in $m_0$, and in fact, they found singularities in figures
plotting the naturalness limits on $m_0$.  These singularities result
from the same numerical `coincidence' responsible for the focus point
mechanism.\footnote{This numerical fact can also be deduced from
earlier papers studying the RG behavior of minimal supergravity. (See,
for example, Ref.~\cite{Alvarez-Gaume:1983gj}.)}  However, although
the sensitivity to $h_t$ was not included in the numerical analysis,
these authors expected a full analysis of naturalness to include this
sensitivity~\cite{Giudice}, and noted that the singularities of the
figures would be eliminated if the sensitivity to $h_t$ were included.
(The sensitivity of the weak scale to the top Yukawa coupling was
considered in more detail in later studies --- see, for example,
Ref.~\cite{Ross:1992tz}.) For this reason, these authors did not claim
that multi-TeV scalars could be natural. Of course, at that time, the
top quark mass was only indirectly bounded, and for most possible
masses, the inclusion of $c_{h_t}$ made no qualitative difference to
the results.

After the discovery of the top quark and the measurement of its mass,
studies of naturalness and the RG properties of minimal supergravity
again found this numerical coincidence (see, e.g.,
Refs.~\cite{Carena:1994bs,Chankowski}).  However, none of these
studies interpreted these results as allowing natural multi-TeV
scalars.  This claim was first made in Ref.~\cite{Feng:2000mn}, where
the issues relevant to the inclusion or exclusion of $c_{h_t}$ were
carefully addressed, the naturalness bounds were investigated
numerically using a full two-loop analysis, and the top mass required
for a weak scale focus point was found to coincide (within
experimental uncertainties) with the measured top mass.  In this
paper, the general requirements for focus point supersymmetry were
also derived, and the potential for this behavior to solve and
ameliorate the supersymmetric flavor and CP problems was noted.

The peculiar value of the top quark mass thus highlights a question,
which, for any other mass, would be of only academic interest: should
the sensitivity to $h_t$ (and other standard model parameters) be
included in calculations of fine-tuning?  Note that whether a
parameter has been measured or not has no bearing on whether its
sensitivity coefficient should be included.  For example, if in the
future the $\mu$ parameter is measured to be $10^{10}~\gev$ to
arbitrarily high accuracy, but our theoretical understanding of
electroweak symmetry breaking has not advanced, $c_{\mu}$ should still
be included in measures of naturalness, and the weak scale should be
considered (highly) fine-tuned.  (Not all naturalness studies take
this view --- see below.)

To address this question, we must first acknowledge the inescapable
model dependence in any naturalness prescription.  In any
supersymmetry study, some fundamental framework must be adopted.  In
studies of other topics, however, there exists, at least in principle,
the possibility of a model-independent study, where no correlations
among parameters are assumed.  This model-independent study is the
most general possible, in that all possible results from any other
(model-dependent) study are a subset of the model-independent study's
results.  In studies of naturalness, however, the correlations
determine the results, and there is no possibility, even in principle,
of a model-independent study in the sense described above.  As an
example, consider a study investigating models where the minimal
supergravity assumptions, in particular, the assumption of scalar
universality, are relaxed.  In such models, the correlations required
by the focus point mechanism are absent.  This study therefore misses
this possibility, and should conclude that it is never possible to
raise all scalar masses far above the TeV level (although the scalar
masses of the first and second generations may be as large as ${\cal
O}(10~\tev)$).

The model dependence of naturalness is present even in the most
general statements concerning fine-tuning.  It is often assumed that,
since the weak scale is $\sim 100~\gev$, supersymmetry parameters of
order 100 GeV will yield a 1 part in 1 fine-tuning in the electroweak
scale.  However, this is at odds with a low energy effective field
theory perspective.  {}From such a point of view, the Higgs mass
receives radiative corrections $\Delta m_h^2 \sim m_{\text{SUSY}}^2 /
16 \pi^2$, so demanding a 1 part in 1 fine-tuning would apparently
allow supersymmetric masses of order $m_{\text{SUSY}} \sim 1~\tev$.
The resolution is that the first statement implicitly assumes a
fundamental theory at some high scale, such as $\mgut$, with
fundamental parameters defined at this high scale.  The radiative
correction is then more precisely $\Delta m_h^2 \sim m_{\text{SUSY}}^2
\ln(\mgut/ \mweak) / 16 \pi^2$, and the large logarithm offsets the
loop factor suppression, yielding a 1 part in 1 fine-tuning for 100
GeV supersymmetry masses.

What about the top quark Yukawa?  {}From a low energy point of view,
one should include all the parameters of the Lagrangian, including
$h_t$.  However, by assuming some underlying high energy motivation by
defining our parameters at $\mgut$, we have already abandoned a purely
low energy perspective.  Once we consider the high energy
possibilities, the case is not so clear. For example, $h_t$ may be
fixed to a specific value (or one of a set of discrete values) in a
sector of the theory unrelated to supersymmetry breaking.  An example
of this is weakly coupled string theory, where $h_t$ may be determined
by the correlator of three string vertex operators and would therefore
be fixed to some discrete value determined by the compactification
geometry.\footnote{Of course, one might argue that string theory may
fix all parameters, including those that break supersymmetry.  Taken
to an extreme, then, no variables are free, and no definition of
naturalness is possible.  Such an approach is equivalent to the
strongest possible anthropic principle, and no more constructive.}  In
such a scenario, it is clearly inappropriate to artificially vary
$h_t$ continuously to determine the sensitivity of the weak scale to
variations in $h_t$. This and other examples leading to the same
conclusion were previously described in Ref.~\cite{Feng:2000zg}.

Clearly, no definitive answer can be given without improved knowledge
of the fundamental theories of flavor and supersymmetry breaking.
Without this knowledge, neither choice is beyond reproach.  However,
given the plausible suggestions from high energy frameworks that the
standard model parameters may be fixed in ways unrelated to
supersymmetry breaking, it is well worth considering the implications
of relaxing the requirement that the weak scale be insensitive to
variations in standard model couplings.  Once we take this approach,
we find it highly suggestive that the measured value of the top quark
mass, along with the simplest of scalar mass boundary conditions, is
exactly what is required to decouple scalars naturally and relieve the
longstanding low energy problems of supersymmetry.

\subsection{Sensitivity vs. Fine-tuning}

The approach of the early papers was criticized in a series of papers
by Anderson and Casta\~{n}o~\cite{AC}.  They pointed out that it is
possible in certain cases that all possible choices of a fundamental
parameter yield large sensitivities.  They argued that in such cases,
only {\em relatively} large sensitivities should be considered
fine-tuned, and drew a distinction between the sensitivity parameters
$c_i$ defined above, and fine-tuning parameters, which they defined as
$\gamma_i \equiv c_i / \bar{c}_i$, with $\bar{c}_i$ an average
sensitivity.  These $\gamma_i$ were then combined to form an overall
fine-tuning parameter.

We agree in principle with these arguments.  In addition to the
virtues noted by Anderson and Casta\~{n}o, the normalization step has the
feature that the fine-tuning is then insensitive to whether the
fundamental parameter is defined to be $m_0$ or $m_0^2$, for example.
However, the averaging procedure may also mask important features. In
their study, Anderson and Casta\~{n}o propose two possible definitions of
$\bar{c}_i$ and show that they yield roughly equivalent results.  One
of these definitions is $\bar{c}_i = \int_{a_i} c_i$, where, as
indicated, the average is taken over a line in parameter space,
varying $a_i$ while holding all other parameters fixed. Adopting this
prescription, in minimal supergravity for a fixed $m_0$, say,
$\gamma_{m_0}$ will be qualitatively the same for top Yukawa couplings
both at the focus point value and far from it: in the latter case, the
sensitivity coefficients $c_{m_0}$ will be much larger, but so will
$\bar{c}_{m_0}$.  We believe this hides a physical effect --- it is
clear that for the focus point top Yukawa coupling, the weak scale is
much less sensitive to variations in $m_0$, and this fact should be
reflected in any definition of naturalness.

To fix this, while preserving the principle virtue of the sensitivity
vs.~fine-tuning distinction, one could define $\bar{c}_i \equiv
\int_{\{m_0, \mgaugino, \ldots, h_t, \ldots\} } c_i$, where one
averages over all of parameter space, including points with different
$h_t$.  This then introduces one overall normalization factor for each
$c_i$, and we have checked that our results are not qualitatively
altered by such a procedure.  It is clear, however, that this modified
Anderson-Casta\~{n}o prescription requires a definition of averaging
region, which introduces additional subjectivity and complications.
Given that our results are not substantially changed, we do not
include this refinement.

\subsection{Naturalness vs. Likelihood}

Finally, an alternative definition of sensitivity coefficient has been
proposed in a series of papers~\cite{RS,Romanino:1999ut}. In these
studies, the definition of Eq.~(\ref{sensitivity}) is replaced by

\begin{equation}
c_i \equiv \left| \frac{\Delta a_i}{m_Z^2} 
\frac{\partial m_Z^2}{\partial a_i} \right| \ ,
\label{alternative}
\end{equation}
where $\Delta a_i$ is the experimentally allowed range of $a_i$. The
intent of this alternative definition is to encode the idea that
naturalness is our attempt to determine which values of parameters are
most likely to be realized in nature.

To contrast this definition with the conventional definition,
consider, for example, the hypothetical scenario described above, in
which our theoretical understanding of supersymmetry has not improved,
but the $\mu$ parameter is measured to be $10^{10}~\gev$ with very
high accuracy. In the standard definition of sensitivity coefficient,
Eq.~(\ref{sensitivity}), the model is fine-tuned.  In our view, this
is as it should be: such a large $\mu$ parameter signals a highly
unnatural situation, and would strongly suggest a deficiency in our
theoretical understanding.  However, by the definition of
Eq.~(\ref{alternative}), $\Delta \mu$ is very small, and so the
electroweak scale is not fine-tuned, even though it is smaller than
$\mu$ by many orders of magnitude.

Naturalness is not simply a measure of our experimental knowledge of
the parameters of nature.  Rather it is a measure of how well a given
theoretical framework explains the parameters realized in nature.  It
is perfectly possible for experimentally likely ranges of parameters
to be unnatural --- this is what the gauge hierarchy and cosmological
constant problems are! --- and to think that this unnaturalness can be
reduced by improved experimental measurements misses this essential
point.

While this is perhaps the most fundamental difference between these
papers and our approach, we conclude with some additional comments
concerning the most recent study of Romanino and
Strumia~\cite{Romanino:1999ut}, as this specifically addresses the
question of the naturalness of multi-TeV scalars.

Naturalness must be calculated in a well-defined framework.  For
example, if we assume minimal supergravity, the focus point works
because the initial value of $m_{H_u}^2$ is $m_0^2$, and the RG
contribution is roughly $-m_0^2$, so the weak scale value vanishes,
independent of $m_0$.  The authors of Ref.~\cite{Romanino:1999ut} ask,
``is a cancellation between [$m_0$] and the radiative contributions to
it more `natural' than a cancellation between different soft terms?''
In our approach, the answer is yes, because in minimal supergravity,
the first two are controlled by the same parameter, whereas different
soft terms cancel only for certain choices of two or more parameters.
Stated in another way, the assumptions of minimal supergravity
guarantee this cancellation just as the assumptions of local quantum
field theory guarantee that the electron's charge is canceled by the
positron's, and it makes no more sense to think of the former
cancellation as fine-tuned than the latter --- it is part of the
assumed framework.\footnote{To be clear, we note that in the case of
minimal supergravity and the focus point, the cancellation is, of
course, not perfect, in contrast to the case of local quantum field
theory and particle/anti-particle charges.  However, for top quark
masses within the current experimental bounds, the cancellation is
complete enough that the sensitivity to multi-TeV $m_0$ is below or of
order the sensitivity to the other ${\cal O}(100~\gev)$ fundamental
parameters, and far below what might naively be expected.}  Of course,
contrived frameworks should be considered less promising, but once the
framework is adopted, one should not vary from its underlying
assumptions.

The authors of Ref.~\cite{Romanino:1999ut} also work in the context of
minimal supergravity.  However, they consider the sensitivity of the
focus point mechanism to ``uncertainties associated with an unknown
sparticle spectrum between 200 GeV and 1 TeV.''  In our approach, the
weak scale threshold corrections are fixed by the input parameters,
and there is no remaining freedom for {\em ad hoc} adjustments of the
sparticle spectrum. All threshold corrections are therefore already
included in our analysis of sensitivity coefficients and in our
results.

Finally, Romanino and Strumia (and others~\cite{Allanach:2000ii})
concentrate their discussion on the case $\tb \approx 10$, which may
leave an impression that the focus point mechanism is operational only
for that specific value of $\tan\beta$. Ref.~\cite{Romanino:1999ut}
also analyzed the effects of uncertainties in $h_t(\mgut)$ on the
focus point scenario.  We reiterate that the RG trajectories of
$m_{H_u}^2$ focus at the weak scale for any value of $\tb \agt 5$.
This was demonstrated numerically in Ref.~\cite{Feng:2000zg}; in the
Appendix, we prove it analytically.  Thus the naturalness of multi-TeV
scalars is guaranteed for virtually all values of $\tb$ allowed by
present constraints on the light Higgs boson mass.

The $\tb$ independence of the focus point is far from trivial.  The
top quark mass and $\tb \agt 5$ fix the top quark Yukawa coupling at
the weak scale.  However, as $\tb$ increases from moderate to large
values, $h_b$ becomes relevant and has two effects: first on the RG
evolution of $h_t$, and, second, directly on the RG trajectories of
the top squark masses and $m^2_{H_u}$.  It is easy to see that these
effects oppose each other.  A non-negligible $h_b$ increases
$h_t(\mgut)$ and the average value of $h_t$ through its RG evolution,
which tends to drive $m_{H_u}^2$ {\em more} negative.  On the other
hand, larger $h_t$ and $h_b$ decrease the average top squark mass,
which pushes $m_{H_u}^2$ {\em less} negative.

What is remarkable, however, is that at one-loop, ignoring negligible
hypercharge effects, these effects {\em exactly} compensate each
other, so that the focus point remains at the weak scale. This is
demonstrated in the Appendix, where we show that the focus point scale
may be written in terms of $h_t(\mweak)$ only, without reference to
$h_t(\mgut)$ or the RG trajectory of $h_b$. Thus, the variation of
$h_t(\mgut)$ is irrelevant to analyses of the focus point: the focus
point mechanism is guaranteed for all $\tb \agt 5$, even though the RG
trajectories of $h_t$, $h_b$ and the third generation squark masses
may vary widely as $\tb$ varies in this range.

\section{Conclusions}
\label{sec:conclusions}

Focus point supersymmetry is motivated by the remarkable `coincidence'
that the precisely measured top quark mass implies that multi-TeV
scalars are natural, given certain simple high scale boundary
conditions.  In this paper, we have considered several
phenomenological consequences of focus point supersymmetry.  We find
that the possibility of {\em all} scalar masses being naturally above
1 TeV has a number of desirable features:

\begin{itemize}

\item The difficulties of many GUT models in accommodating both gauge
coupling unification and proton decay constraints are reduced.

\item Constraints from EDM measurements on unknown phases are less
stringent by one to two orders of magnitude, and current constraints
may be satisfied with ${\cal O}(0.1)$ phases.

\item The Higgs boson mass is predicted to be at or above 115 GeV in
focus point scenarios, consistent with current constraints and the
recent evidence for a 115 GeV Higgs boson at LEP.  Such large masses
are typically difficult to obtain without fine-tuning, but are
achieved naturally in focus point supersymmetry through heavy top and
bottom squarks, without the need to appeal to large CP violating
phases or heavy gauginos.

\end{itemize}
In addition, we analyzed the implications of focus point supersymmetry
for two other important constraints on supersymmetric theories, the
muon MDM and $\bsg$.  In particular, for the muon MDM, while an
observable deviation from the standard model is consistent with focus
point supersymmetry with high $\tb$, a near future measurement
consistent with the standard model will exclude conventional regions
of minimal supergravity parameter space, and will strongly prefer
focus point scenarios.

Finally, we have concluded this study with an extended discussion of
various naturalness prescriptions.  We have identified and highlighted
a number of key differences between our prescription and others in the
literature.  By far the most ambiguous and important issue, in our
view, is the question of whether one should include sensitivities to
standard model couplings in attempts to quantify the success of
supersymmetry in solving the gauge hierarchy problem.  We have
identified several scenarios in which such sensitivities should not be
included.  Perhaps most suggestive, however, is the fact that by
excluding the sensitivity to standard model parameters, the measured
top quark mass implies that multi-TeV scalars are natural for the
simplest possible boundary condition of universal scalar masses.  If
this is more than a coincidence, the top quark mass is our hint that
the low energy problems of supersymmetry are but a mirage, and the
mass scale of all squarks and sleptons actually lies well above a TeV.

\section*{Acknowledgments}

We are grateful to C.~Kolda and T.~Moroi for collaboration in the
early stages of this work and to F.~Wilczek for fruitful discussions.
We also thank G.~Anderson, P.~Chankowski, A.~Romanino, and A.~Strumia
for extensive correspondence and conversations. This work was
supported in part by the U.~S.~Department of Energy under contracts
DE--FG02--90ER40542 and DE--AC02--76CH03000 and in part by funds
provided by the U.~S.~Department of Energy under cooperative research
agreement DF--FC02--94ER40818.  J.L.F. acknowledges with gratitude the
support of a Frank and Peggy Taplin Membership.

\appendix

\section*{Dependence of Focus Point on Yukawa Couplings}
\label{app:dependence}

In this Appendix, we show that in minimal supergravity, the focus
point scale is determined only by the gauge couplings and the weak
scale value of the top Yukawa coupling $h_t$.  More precisely, we show
that the renormalization scale at which the $m_{H_u}^2$ contours meet
may be written only in terms of $h_t$ at the weak scale, with no
reference to the rest of the top quark Yukawa RG trajectory (e.g., its
value at $\mgut$) or to the bottom Yukawa coupling $h_b$.  This
demonstrates that if the focus point is at the weak scale for, say,
$\tb =5$, it remains there for all $\tb>5$.  In
Ref.~\cite{Feng:2000zg} this was shown analytically for $h_b \ll h_t$
(moderate $\tb$) and $h_b=h_t$ (high $\tb$), and also numerically for
all $\tb$.  Here we demonstrate this analytically for all $\tb$,
neglecting the tau Yukawa coupling, but making no assumptions about
the relative magnitudes of $h_t$ and $h_b$.  An abbreviated version of
this proof was presented in Ref.~\cite{Feng:1999sw}.

To analyze the focus point, it is convenient to define
\begin{eqnarray}
t        &\equiv& {1\over 2\pi} \ln \left({Q\over\mgut}\right)\ , \\
\alpha_i &\equiv& \frac{g_i^2}{4 \pi}\ , \\
Y_i      &\equiv& \frac{h_i^2}{4 \pi}\ , \\
m_i^2    &\equiv& m_i^2|_{\rm p} + \Delta_i^2 \label{notation} \ ,
\end{eqnarray}
where $Q$ is the renormalization scale,\footnote{Notice that we have
rescaled the variable $t$ relative to its conventional definition, in
order to simplify the equations to follow.} $g_i$ and $h_i$ are gauge
and Yukawa couplings, respectively, and $m_i^2$ are scalar masses.
Following the notation of
Refs.~\cite{Feng:2000hg,Feng:2000mn,Feng:2000zg}, we separate the
scalar mass into $m_i^2|_{\rm p}$, a particular solution to the RG
equations, and $\Delta_i^2$, the remaining homogeneous part.

We now keep only the top and bottom Yukawa couplings, and neglect the
small hypercharge difference in the $Y_t$ and $Y_b$ RG equations.
With these approximations, the one-loop RG equations for the couplings
are

\begin{eqnarray}
\dot{\alpha}_3 &=& - 3          \alpha_3^2 \ , \qquad
\dot{\alpha}_2  =               \alpha_2^2 \ , \qquad
\dot{\alpha}_1  =  \frac{33}{5} \alpha_1^2 \ , \\
\dot{Y}_t      &=& Y_t\ [6Y_t+ Y_b - r(\alpha)]\ , \label{YtRGE} \\
\dot{Y}_b      &=& Y_b\ [ Y_t+6Y_b - r(\alpha)]\ , \label{YbRGE} 
\label{YRGEs}
\end{eqnarray}
where $\dot{}\equiv d/dt$ and $r(\alpha)\equiv\frac{16}{3} \alpha_3 +
3 \alpha_2 + \frac{13}{15} \alpha_1$ is a function of gauge couplings
only.  The homogeneous scalar mass evolution is given by

\begin{equation}
\dot{\bold{\Delta}}^2 = \bold{N} \bold{\Delta}^2 \ ,
\label{DeltaRGEs}
\end{equation}
where

\begin{equation}
\bold{N} = \left[ \begin{array}{ccccc}
        3Y_t & 3Y_t & 3Y_t & 0 & 0 \\
        2Y_t & 2Y_t & 2Y_t & 0 & 0 \\
        Y_t & Y_t & Y_t + Y_b & Y_b& Y_b \\
        0 & 0 & 2Y_b & 2Y_b & 2Y_b \\
        0 & 0 & 3Y_b & 3Y_b & 3Y_b
        \end{array} \right] \ ,
\end{equation}
and $\bold{\Delta}^2 = \left[ \Delta_{H_u}^2, \Delta_{U_3}^2,
\Delta_{Q_3}^2, \Delta_{D_3}^2, \Delta_{H_d}^2 \right]^T$, with $U_3$,
$Q_3$, and $D_3$ the third generation squark multiplets, and $H_u$ and
$H_d$ the up- and down-type Higgs multiplets.

Equation~(\ref{DeltaRGEs}) is a set of five coupled differential
equations, but the simple form of $\bold{N}$ implies that the RG
evolution of three degrees of freedom is trivial.  To make this
explicit, define

\begin{equation}
\frac{\bold{\Delta}^2 (t)}{m_0^2} \equiv 
c_1(t) \left[ \begin{array}{c}
        3 \\ 2 \\ 1 \\ 0  \\ 0
        \end{array} \right]
+ c_2(t) \left[ \begin{array}{c}
        0 \\ 0 \\ 1 \\ 2 \\ 3
        \end{array} \right]
+ c_3(t) \left[ \begin{array}{r}
        1 \\ -1 \\ 0 \\ 0 \\ 0
        \end{array} \right]
+ c_4(t) \left[ \begin{array}{r}
        0 \\ 1 \\ -1 \\ 1 \\ 0
        \end{array} \right]
+ c_5(t) \left[ \begin{array}{r}
        0 \\ 0 \\ 0 \\ -1 \\ 1
        \end{array} \right] ,
\end{equation}
where we have factored out an overall mass scale $m_0$ to make the
$c_i$ dimensionless.  Equation~(\ref{DeltaRGEs}) then reduces to

\begin{eqnarray}
\dot{c}_1 &=& Y_t(6c_1+c_2)\ , \nonumber \\
\dot{c}_2 &=& Y_b(c_1+6c_2)\ , 
\label{cRGEs}
\end{eqnarray}
and $\dot{c}_3 = \dot{c}_4 = \dot{c}_5 = 0$.  

We now solve these equations in full generality.
Equations~(\ref{cRGEs}) form a linear homogeneous system of
first-order ordinary differential equations with variable
coefficients.  No general method of solution exists for such
systems~\cite{Kamke}. However, in this case, the variable coefficients
$Y_t$ and $Y_b$ satisfy Eqs.~(\ref{YtRGE}) and (\ref{YbRGE}), and this
allows us to integrate these equations after a well-chosen ansatz for
the form of the $c_i$.

Let us make a change of variables \cite{Feng:1999sw}
\begin{eqnarray}
c_1(t)   &=& c_1^0 + Y_t(t) p(t) \ ,  \label{ctansatz}\\
c_2(t)   &=& c_2^0 + Y_b(t) q(t) \ ,  \label{cbansatz}
\end{eqnarray}
where $c_i^0 \equiv c_i(0)$ and the boundary condition for the new
variables $p$ and $q$ is $p(0)=q(0)=0$.  Substituting these forms for
$c_1$ and $c_2$ into Eqs.~(\ref{cRGEs}) and using the Yukawa RG
Eqs.~(\ref{YtRGE}) and (\ref{YbRGE}), we find
\begin{eqnarray}
\dot{p}&=&   Y_b (q-p) + rp + 6 c_1^0 + c_2^0 \label{dpdt} \ , \\
\dot{q}&=&   Y_t (p-q) + rq + 6 c_2^0 + c_1^0 \label{dqdt} \ .
\end{eqnarray}
The difference of Eqs.~(\ref{dpdt}) and (\ref{dqdt}) yields a simple
first order linear inhomogeneous differential equation for $p-q$
\begin{equation}
{d\over dt}(p-q)\ =\ -\ (Y_t+Y_b-r)(p-q) + 5 (c_1^0 -c_2^0) \ ,
\label{dpmqdt}
\end{equation}
which integrates to \cite{Feng:1999sw}
\begin{equation}
p-q = 5 (c_1^0-c_2^0) e^{- \int (Y_t + Y_b -r)}
\int e^{\int (Y_t + Y_b -r)} \ .  \label{pmq}
\end{equation}

To solve for $p$ or $q$, substitute Eq.~(\ref{pmq}) into
Eq.~(\ref{dpdt}) or Eq.~(\ref{dqdt}), respectively.  The resulting
differential equation is again easily solved, and the solution for $p$
is
\begin{equation}
p(t)= (6c_1^0+c_2^0) e^{\int r}\int e^{-\int r}
- 5(c_1^0 - c_2^0)e^{\int r}
\int  \left[ Y_b e^{- \int (Y_t + Y_b)}
\int e^{\int (Y_t + Y_b - r)} \right] \ .
\end{equation}
The final solution for $c_1(t)$ is then \cite{Feng:1999sw}
\begin{eqnarray}
c_1(t)&=& c_1^0 + Y_t(t) e^{ \int_0^t dt_1 r(t_1)}\Biggl\{
(6c_1^0 + c_2^0)\int_0^t dt_1 e^{-\int_0^{t_1}dt_2 r(t_2)} \nonumber \\
    &-& 5(c_1^0 - c_2^0)
\int_0^t dt_1  Y_b(t_1) e^{-\int_0^{t_1} dt_2 [Y_t(t_2)+Y_b(t_2)]}
\int_0^{t_1} dt_2 e^{\int_0^{t_2}dt_3 [Y_t(t_3) + Y_b(t_3) - r(t_3)]}
\Biggr\} \ , \label{finalc1}
\end{eqnarray}
and $c_2(t)$ is obtained by interchanging $t\leftrightarrow b$, and $1
\leftrightarrow 2$.

The focus point scale $t_F$ is given by

\begin{equation}
\Delta_{H_u}^2(t_F) = c_3 + 3 c_1(t_F) = 0 \ .
\end{equation}
In the case of a universal scalar mass, the initial conditions are
$c_i(0) = \left[ 3/7, 3/7, -2/7, -1/7, -2/7 \right]$.
Equation~(\ref{finalc1}) then becomes

\begin{equation}
c_1(t) = \frac{3}{7} + 3 Y_t(t) e^{ \int_0^t dt_1 r(t_1)}
\int_0^t dt_1 e^{-\int_0^{t_1}dt_2 r(t_2)} \ .
\end{equation}
Note the great simplification following from $c_1^0 = c_2^0$.  The
focus point is therefore fixed by the constraint

\begin{equation}
Y_t(t_F) e^{ \int_0^{t_F} dt_1 r(t_1)}
\int_0^{t_F} dt_1 e^{-\int_0^{t_1}dt_2 r(t_2)} = - \frac{1}{9} \ .
\label{mtcondition}
\end{equation}

We see that $t_F$ depends on the entire RG trajectories of the gauge
couplings and on $Y_t$ at the focus point, but {\em is independent of
the rest of the $Y_t$ trajectory and is also entirely independent of
$Y_b$}.  For the physical top mass $\mt \approx 174~\gev$ and $\tb
\approx 5$, we know that the the focus point is at the weak scale
\cite{Feng:2000mn}.  As we raise $\tb$, $Y_t(t_F) =
Y_t(t_{\text{Weak}})$ remains approximately constant to reproduce the
physical top quark mass, but $Y_b$ increases.  Eventually, $Y_b$ will
be large and the RG trajectory of $Y_t$ (and, of course, $Y_t(\mgut)$)
will be modified accordingly.  The scalar mass RG trajectories are
then also modified.  Remarkably, the analysis above shows that despite
this, the focus point of the $m_{H_u}^2$ trajectories remains at the
weak scale.  The focus point therefore remains at the weak scale for
all $\tb \agt 5$, and, in particular, is independent of the GUT scale
value of $Y_t$, as long as $Y_t$ at the weak scale remains fixed, as
it must to be consistent with the measured top quark mass.

With the analytic solution at hand, it is now straightforward to
generalize the focus point discussion to the case of scalar mass
non-universality.  Using the empirical relation
Eq.~(\ref{mtcondition}), we can write the focus point condition in the
form
\begin{equation}
3c_1^0 - c_2^0 + 3c_3 = 0\ .
\label{FPcondition}
\end{equation}
Any set of non-universal boundary conditions satisfying
Eq.~(\ref{FPcondition}) will exhibit a focus point at the weak scale,
at least for a certain range of (moderate) values of $\tb$.
Furthermore, if in addition
\begin{equation}
c_1^0 - c_2^0 = 0\ ,
\label{tanbcondition}
\end{equation}
then a weak scale focus point exists for {\em any} value of
$\tan\beta\agt5$. The most general set of non-universal scalar
boundary conditions satisfying both Eqs.~(\ref{FPcondition}) and
(\ref{tanbcondition}) is~\cite{Feng:2000zg}
\begin{equation}
\left[ \begin{array}{c}
m_{H_u}^2\\ m_{U_3}^2\\ m_{Q_3}^2\\ m_{D_3}^2\\ m_{H_d}^2
\end{array} \right]
\ =\ m_0^2\ 
\left[ \begin{array}{lll}
1&    &     \\
1&+\ x&     \\
1&-\ x&     \\
1&+\ x&-\ x'\\
1&    &+\ x'
\end{array} \right]
\end{equation}
with both $x$ and $x'$ arbitrary.

\end{document}